\documentclass[
reprint, onlinecite, superscriptaddress,
amsmath,amssymb,
aps,prl
]{revtex4-2}

\usepackage{makecell}
\usepackage{graphicx}
\usepackage{dcolumn}
\usepackage{bm}
\usepackage{longtable}
\usepackage{physics}
\usepackage{amsmath}
\usepackage{multirow}
\usepackage{colortbl}
\usepackage{booktabs}
\usepackage[
    colorlinks=true,
    linkcolor=blue,
    citecolor=blue,
    urlcolor=blue
]{hyperref}

\begin{document}
\title{Two-dimensional vertically polarized Hg$_{3}$AsSe$_{4}$I monolayer for efficient photocatalytic water-splitting: promoting carrier separation by intrinsic electric field and Rashba effect}

\author{Xinfeng Chen}
\altaffiliation{These authors contributed equally to this work.}
\affiliation{Frontier Institute of Science and Technology, State Key Laboratory of Electrical Insulation and Power Equipment, Xi’an Jiaotong University, Xi’ an 710049, People's Republic of China}
\affiliation{Key Lab of advanced optoelectronic quantum architecture and measurement (MOE), Beijing Key Laboratory of Quantum Matter State Control and Ultra-Precision Measurement Technology, and School of Physics, Beijing Institute of Technology, Beijing 100081, China}

\author{Wenchao Shan}
\altaffiliation{These authors contributed equally to this work.}

\author{Fengfeng Ye}

\author{Gaoyang Gou}
\email{gougaoyang@mail.xjtu.edu.cn}
\affiliation{Frontier Institute of Science and Technology, State Key Laboratory of Electrical Insulation and Power Equipment, Xi’an Jiaotong University, Xi’ an 710049, People's Republic of China}

\begin{abstract}  

Efficient separation of photo-excited electron-hole pairs is essential for developing the high-performance photocatalysts towards light-driven water-splitting applications. To this end, photocatalytic performances of two-dimensional (2D) semiconducting ferroelectric (FE) materials with out-of-plane polarizations have been extensively explored. However, out-of-plane polarizations in 2D FE materials are susceptible to the critical thickness limitation and can be easily compensated by surface adsorbates. On the other hand, 2D vertically polarized materials with stable and irreversible out-of-plane polarizations may overcome the critical thickness limitation, enabling the practical advantage for spatial separation of photo-excited electron-hole pairs during the photocatalytic reactions. In the current work, 2D vertically polarized Hg$_{3}$AsSe$_{4}$I, an experimentally synthesized van der Waals (vdW) layered material, has been systematically investigated as a high-performance 2D photocatalyst. Owing to its semiconducting band gap suitable for visible-light absorption, high carrier mobility, and desirable band edge alignment ideally matching water reduction and oxidation potentials, Hg$_{3}$AsSe$_{4}$I monolayer fulfills both optical and electronic prerequisites for photocatalytic water-splitting reactions. Besides the stable vertical polarization able to persist in Hg$_{3}$AsSe$_{4}$I monolayer, the dual mechanism for efficient separation of photo-excited carriers has also been demonstrated. Rashba spin-orbit coupling (SOC) of large strength emerges within 2D Hg$_{3}$AsSe$_{4}$I, splitting the band edges into spin-resolved band branches with unique spin-momentum locking characters. The resulting spin-selection rules can effectively prevent the direct recombination of electrons and holes around the band extrema, further extending the effective lifetime of photo-excited carriers. In fact, the long carrier lifetime up to 21.57 ns has been predicted for Hg$_{3}$AsSe$_{4}$I monolayer by time-domain nonadiabatic molecular dynamics (NAMD) simulations. Our work demonstrates the synergy of stable out-of-plane polarization and Rashba effect in promoting the efficient carrier separation, and also provides a rational guideline for exploring 2D vertically polarized materials for photocatalytic applications.

\end{abstract}

\maketitle


\section{\label{sec:level1}I.INTRODUCTION}

Sunlight-driven photocatalytic water-splitting reaction manifests as a practical method to massively convert the abundant solar energy into clean hydrogen (H$_{2}$) fuel\cite{000343993100005}, offering a viable solution to address the environmental pollution and energy crises caused by excessive fossil consumption\cite{000446615700012,000513814400011,000282017400007}. Ever since the landmark discovery of TiO$_{2}$ as a photocatalyst for water-splitting reaction\cite{A1972M852600031}, tremendous research efforts have been devoted to explore other efficient photocatalysts(e.g., metal oxides\cite{000312552900035}, nitrides\cite{000463032200071}, sulfides\cite{000865240400001}) for water-splitting reactions. However, most three-dimensional (3D) bulk photocatalysts have the intrinsic limitations, such as poor absorption of incident sunlight, low carrier mobilities, easy recombination of photo-excited electron-hole pairs and limited surface areas unable to provide the abundant active reaction sites, which greatly hinder their practical photocatalytic efficiency\cite{000440682100058}.

Thanks to the rising of 2D vdW layered materials and productive research progress made in exploration of 2D materials for photocatalytic applications\cite{000367280100002,000414506400026}, 2D photocatalytic materials are found to characterize the unique advantages absent in traditional bulk photocatalysts. Specifically, the electronic properties and especially energy band gaps ($E_{g}$) for most semiconducting 2D materials are readily tunable via strain\cite{000320485100100,000435525100068,000772697600001}, chemical functionalization\cite{000428509000001}, and multilayer stacking\cite{000526353000046}, rendering the capacity for efficient absorption of incident sunlight. 2D materials with high crystal quality and atomic-layer thickness usually have ultrahigh carrier mobilities far superior than 3D bulk photocatalysts\cite{000936535100002,000704372200007}. Moreover, the large surface volume ratio of 2D layered materials can guarantee the abundant photocatalytic active sites to participate water-splitting reactions\cite{000348922100002}. Up to date, tremendous 2D photocatalysts, such as graphene oxides\cite{000327749400005}, transition metal dichalcogenides(e.g., MoS$_{2}$ and WS$_{2}$)\cite{000704372200007,000343026700030,000537317200001} and g-C$_{3}$N$_{4}$\cite{001633482900004}, have been extensively investigated and successfully applied for photocatalytic water-splitting reactions. However, there also exists a fundamental disadvantage for 2D layered photocatalysts: owing to the reduced system dimension and enhanced Coulomb screening effect, photo-excited electrons and holes in 2D materials are more inclined to recombine and form the stable excitons with large binding energies, rather than participating the water reduction and oxidation reactions. Therefore, the improvement of photocatalytic efficiency of 2D materials calls for the effective strategy for preventing photo-excited electron-hole pairs from recombination.

To address the above mentioned disadvantages, a new mechanism have been proposed for optimizing photocatalytic efficiency of 2D photocatalysts: taking advantage of spontaneous out-of-plane polarizations appeared in 2D FE and polar materials, the associated internal electric fields are able to effectively separate photo-exicted electrons and holes from recombinations\cite{000509425600047,000875690400001}, as well as regulate the alignment of valence and conduction band edges with respect to the water reduction and oxidation potentials\cite{000331938800014}. To this end, photocatalytic performance of various 2D FE materials (e.g., CuInP$_{2}$S$_{6}$\cite{000679917500073}, $\alpha$-In$_{2}$Se$_{3}$\cite{000440682100058}), Janus transition metal dichalcogenides\cite{000423981200044,000455124900015,001613285000005,000546698600042,001570579100003} and 2D heterostructures with either Type-II band alignment or Z-scheme photocatalytic mechanism are under extensive investigations\cite{000722893000001,001724482600001,001500375600002,001416048100007}. However, the spontaneous out-of-plane polarizations in 2D FE materials usually suffer from the ``critical thickness'' limitation\cite{000221934300044,000470149000044}, room temperature stable out-of-plane ferroelectricity that can persist at the monolayer thickness remains quite rare in experiments. Moreover, during water-splitting reactions, out-of-plane polarizations of 2D FE materials can be suppressed or completely compensated by surface adsorbates\cite{001670039100003}. On the other hand, the preparation of Janus monolayers or 2D heterostructures requires complicated fabrication techniques\cite{000406868800011}. Massively producing 2D Janus or heterostructured materials for photocatalytic applications is not feasible in experiments. Therefore, searching for the alternative single-phase 2D materials with easy experimental synthetic approach, room temperature stable out-of-plane polarization able to persist at ultra-thin atomic thickness, and optimal photocatalytic performance is highly desirable.
 
In the current work, based on first-principles calculations and time-domain NAMD simulations, we uncover the experimentally synthesizable 2D vertically polarized Hg$_{3}$AsSe$_{4}$I monolayer as an efficient photocatalyst whose photocatalytic activity remains unexplored yet. As a semiconductor with suitable $E_{g}$, high carrier mobility and ideal alignment of conduction and valence band edges with respect to water reduction and oxidation potentials, Hg$_{3}$AsSe$_{4}$I monolayer meets both optical and electronic prerequisites as an efficient 2D photocatalyst. In particular, stable and irreversible out-of-plane polarization can persist in Hg$_{3}$AsSe$_{4}$I monolayer, leading to the internal electric field for spatial separation of photo-excited electron-hole pairs. With breaking of spatial inversion symmetry along the out-of-plane direction and significant relativistic effects induced by heavy elements, Rashba SOC emerges in Hg$_{3}$AsSe$_{4}$I monolayer, imposing additional spin-selection rules that forbid the direct recombination of electrons and holes around the band edges. The dual mechanism associated with out-of-plane polarization and Rashba effect for efficient separation of photo-excited carriers within Hg$_{3}$AsSe$_{4}$I monolayer is therefore established. Lastly, the ultra-long carrier lifetime within Hg$_{3}$AsSe$_{4}$I monolayer is validated by time-domain NAMD simulations.

\section{\label{sec:level1}II.COMPUTATIONAL METHODS}
\subsection{\label{sec:level2}First-principles Calculations}

Our first-principles calculations are performed based on density functional theory (DFT) methods as implemented in the Vienna $Ab-initio$ Simulation Package (VASP)\cite{1996VT67500040,1996VF38900003}. We adopt the Perdew–Burke–Ernzerhof (PBE) exchange-correlation functional within the generalized gradient approximation (GGA)\cite{1996VP22500044}. A plane-wave basis set within the projector augmented wave (PAW) method\cite{1994QB02200016} is employed, using an energy cutoff of 600 eV. The interlayer vdW interactions within 2D layered Hg$_{3}$AsSe$_{4}$I are described using DFT-D3 method developed by Grimme\cite{000276971500005}. The crystallographic $a$, $b$ and $c$ axes refer to two planar and one vertical directions of layered Hg$_{3}$AsSe$_{4}$I, respectively. 2D Hg$_{3}$AsSe$_{4}$I layers are simulated as slabs containing at least 20 \AA $ $ vacuum along $c$ axis. The Brillouin zones of Hg$_{3}$AsSe$_{4}$I 3D bulk and 2D layers are sampled based on Monkhorst-Pack scheme\cite{1976BV08800009}, using 12$\times$12$\times$12 and 12$\times$12$\times$1 $k$-point grids, respectively. The dipole correction is included in the vacuum region of 2D Hg$_{3}$AsSe$_{4}$I layers\cite{000080570800036}. The atomic positions and lattice parameters are fully optimized until the residual Hellmann-Feynman forces and stress are less than 0.001 eV/\AA $ $ and 0.1 kbar, respectively. A finite difference method is applied to calculate the phonon spectrum of Hg$_{3}$AsSe$_{4}$I\cite{A1997XA46500021}. The Heyd-Scuseria-Ernzerhof (HSE06) hybrid functional is employed to calculate the band structure and optical absorption properties\cite{000238758700058}. SOC effect is self-consistently included during the electronic structure calculations.

\subsection{\label{sec:level2}Nonadiabatic Molecular Dynamics Simulations}

The dynamic properties of photogenerated carriers inside layered Hg$_{3}$AsSe$_{4}$I are simulated using nonadiabatic molecular dynamics (NAMD) simulations within time-domain DFT (TDDFT) implemented in Hefei-NAMD code\cite{000491351500006}. TDDFT extends the framework of density functional theory (DFT) from the static properties to the time-dependent properties simulations. The Kohn-Sham (KS) approach of TDDFT maps an interacting many-body system into a non-interacting particle system, supposing two systems with identical electron densities. The time-dependent charge density of the interacting system can thus be obtained from the time-dependent KS orbitals:

\begin{equation}\label{Eq1}
  n(\textbf{r},t) = \sum_{i = 1}^{N}|\psi_{i}(\textbf{r},t)|^{2}
\end{equation}

\noindent where $\psi_{i}(\textbf{r},t)$ represents the single-electron KS orbitals, $N$ stands for the number of electrons occupying the orbitals. By employing the time-dependent variational principle, we can derive a set of single-particle equations for the evolution of KS orbitals:

\begin{equation}\label{Eq2}
  i\hbar\frac{\partial}{\partial t}\psi_{i}(\textbf{r},t) = H(\textbf{r},\textbf{R},t) \psi_{i}(\textbf{r},t)
\end{equation}

The expansion of single-electron orbitals in the adiabatic KS orbital basis is given by

\begin{equation}\label{Eq3}
  \psi_{i}(\textbf{r},t) = \sum_{k} c_{ik}(t)|\varphi_{k}(\textbf{r};\textbf{R})\rangle
\end{equation}

Combining Eq.~(\ref{Eq2}) and Eq.~(\ref{Eq3}), one can obtain equations for the expanding coefficients:

\begin{equation}\label{Eq4}
  i\hbar\frac{\partial}{\partial t}c_{j}(t) = \sum_{k} c_{ik}(t)(\varepsilon_{k}\delta_{jk}+d_{jk})
\end{equation}

where $\varepsilon_{k}$ represents the energy eigenvalues of adiabatic KS state $k$. The nonadiabatic coupling can be expressed as follows:
\begin{align}
\label{Eq5}
  d_{jk} &= -i\hbar \langle \varphi_{j}(\textbf{r};\textbf{R})|\nabla_{\textbf{R}}\varphi_{k}(\textbf{r};\textbf{R})\rangle \cdot \frac{d\textbf{R}}{dt} \\
         &= -i\hbar \langle \varphi_{j}(\textbf{r};\textbf{R})|\frac{\partial\varphi_{k}(\textbf{r};\textbf{R})}{\partial t}\rangle
\end{align}

The $Ab$ $initio$ molecular dynamics (AIMD) simulations are performed on 2D Hg$_{3}$AsSe$_{4}$I monolayer configuration around 300 K\cite{000413057500080} for 2 ps with a time step of 1.0 fs. An adiabatic MD microcanonical trajectory of 5 ps is then generated. The corresponding wave function is also produced. The dynamic evolution of photogenerated carriers are statistically acquired by averaging 100 different initial configurations\cite{000629172200020} and sampling 2 $\times$ 103 trajectories for each MD trajectory\cite{000679917500073}. The quantum-classical decoherence-induced surface hopping (DISH) method\cite{000083345100013} is employed to predict the probability for hopping between interacting states based on the evolution of adiabatic basis coefficients.


\section{\label{sec:level1}III. RESULTS AND DISCUSSION}

\begin{figure*}[t]
\centering
\includegraphics[width=0.75\textwidth]{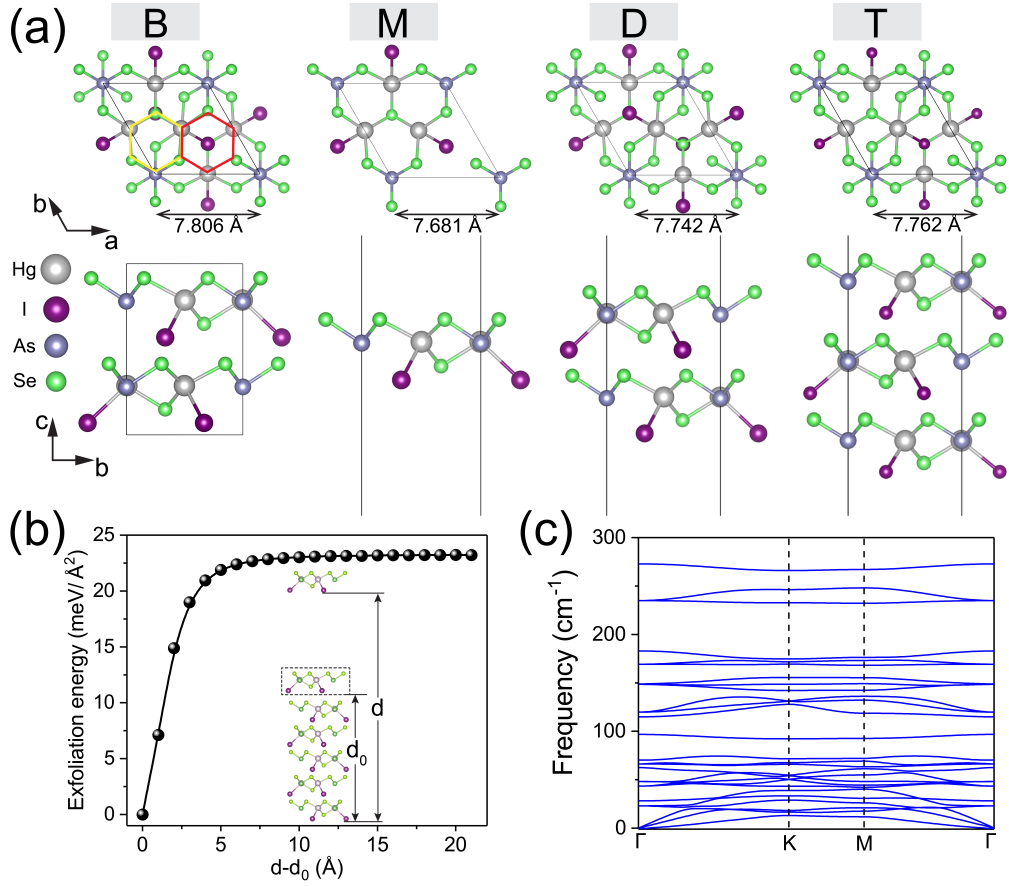}
\vspace{-3pt}
\caption{(a) The top and side views of the atomic structures of Hg$_{3}$AsSe$_{4}$I in bulk (B), monolayer (M), double-layer (D) and trilayer (T) structural forms. Hg, As, Se and I atoms are marked by silver, gray, green and violet, respectively. (b) The variation of the system energy as a function of the interlayer spacing, when Hg$_{3}$AsSe$_{4}$I monolayer is exfoliated from the bulk phase. (c) Our simulated phonon spectrum for Hg$_{3}$AsSe$_{4}$I monolayer, which is free of imaginary frequency.}
\label{Figure 1}
\end{figure*}

In experiment, single crystal Hg$_{3}$AsSe$_{4}$I in bulk structural form has been prepared via the chemical vapor transport reaction method\cite{000165858100035}. FIG.~\ref{Figure 1}(a) displays the atomic structure of layered Hg$_{3}$AsSe$_{4}$I bulk, which crystallizes in the acentric $P6_{3}mc$ space group of hexagonal symmetry and contains two periodically stacked Hg$_{3}$AsSe$_{4}$I monolayers with $AB$ stacking sequence. Each Hg$_{3}$AsSe$_{4}$I monolayer is composed of six-membered Hg$_{2}$AsSe$_{3}$ and Hg$_{3}$Se$_{3}$ rings interconnected along the two planar directions (indicated by yellow and red hexagons in FIG.~\ref{Figure 1}(a)). Along the out-of-plane direction, Se anions are located at the top surface, while Se and I anions coexist at the bottom surface of Hg$_{3}$AsSe$_{4}$I monolayer, rendering the asymmetric anionic distributions between top and bottom surfaces for each monolayer. As a result, Hg$_{3}$AsSe$_{4}$I monolayer characterizes a net electric dipole, and an irreversible out-of-plane polarization of 7.5 $\mu$C/cm$^{2}$ is accumulated and predicted for the Hg$_{3}$AsSe$_{4}$I bulk phase\cite{001291654100001}. As a vdW layered 2D material with weak interlayer binding energy ($E_b$ = -23.85 meV/{\AA}$^{2}$, illustrated in FIG.~\ref{Figure 1}(b)), the exfoliation of 2D Hg$_{3}$AsSe$_{4}$I layers (e.g., monolayer (M), double-bilayer (D) and trilayer (T) displayed in FIG.~\ref{Figure 1}(a)) from the bulk (B) phase can be expected in experiment. After exfoliation, 2D Hg$_{3}$AsSe$_{4}$I layers adopt the same in-plane structural framework as the bulk crystal, except for the contraction of their in-plane lattice parameters by a small magnitude. The dynamic stabilities of 2D Hg$_{3}$AsSe$_{4}$I layers are further examined, even Hg$_{3}$AsSe$_{4}$I monolayer is vibrationally stable and free of structural instability with imaginary frequency (see phonon spectrum in FIG.~\ref{Figure 1}(c)). Moreover, owing to the interruption of out-of-plane periodicity, 2D vertically polarized Hg$_{3}$AsSe$_{4}$I layers possess two nonequivalent surfaces (i.e., top and bottom surfaces) that can favorably interact with incoming H$_{2}$O molecules during the water-splitting reaction (see the simulated H$_{2}$O adsorption configurations on the top and bottom surfaces of 2D Hg$_{3}$AsSe$_{4}$I in FIG. S1 and FIG. S2 in the Supplemental Material\cite{support}). 


\begin{figure*}[t]
\centering
\includegraphics[width=0.97\linewidth]{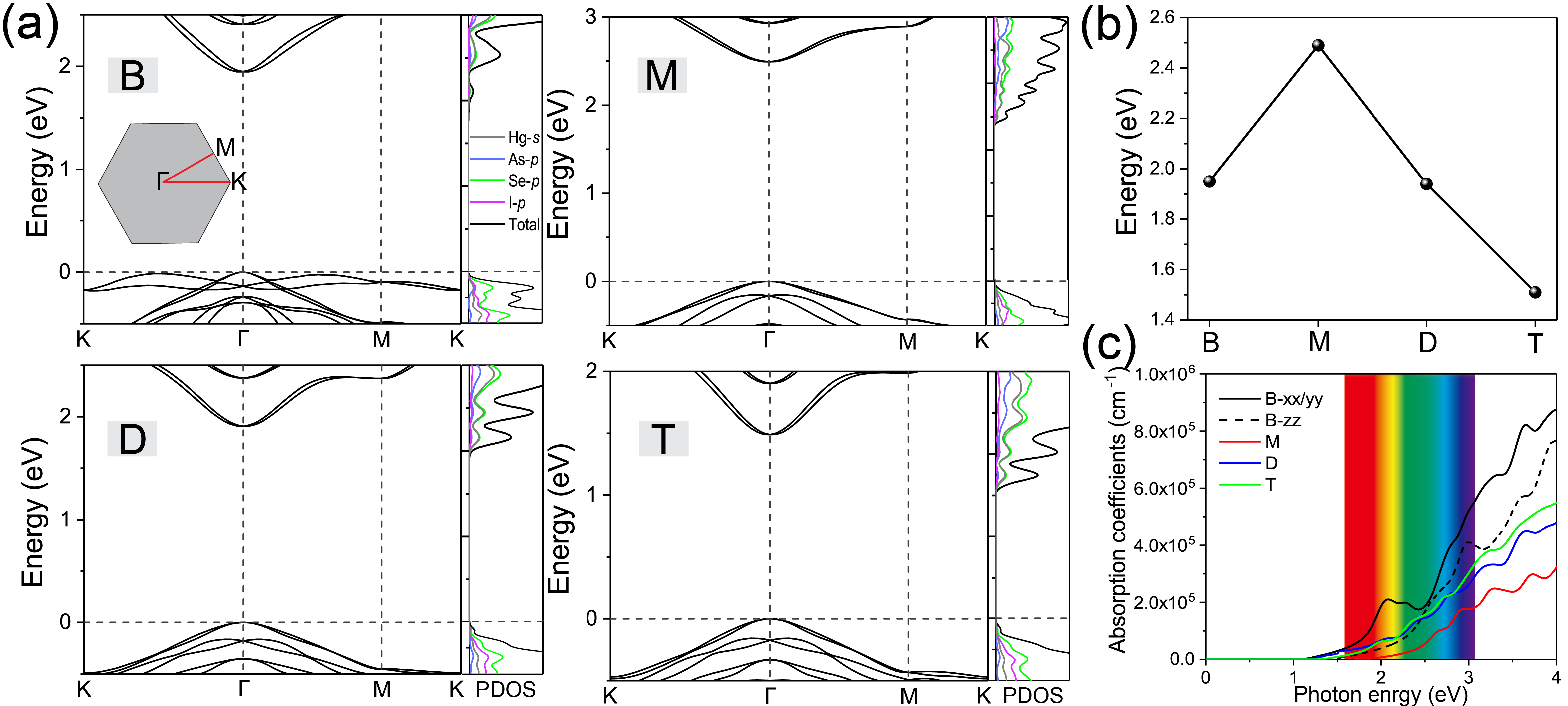}\vspace{-3pt}
\caption{(a) The calculated energy band structures and projected density of states (PDOS) of Hg$_{3}$AsSe$_{4}$I in B, M, D, and T structural forms, using the HSE06 hybrid functional with SOC effect self-consistently included. Rashba effects manifest as the splitting of valence and conduction band edges into the spin-split band branches. The insert shows the irreducible Brillouin zone for hexagonal Hg$_{3}$AsSe$_{4}$I layers. (b) The simulated energy band gaps $E_{g}$ and (c) optical absorption coefficients $\eta_{\omega}$ of Hg$_{3}$AsSe$_{4}$I in B, M, D, and T structural forms. The visible light of solar spectrum is marked by the iridescent color.}
\label{Figure 2}
\end{figure*}

The photocatalytic water-splitting reaction requires the efficient absorption of incident sunlight by photocatalysts. To this end, the electronic and optical absorption properties of 2D layered Hg$_{3}$AsSe$_{4}$I are under simulations. FIG.~\ref{Figure 2}(a) displays the simulated energy band structures and projected density of states (PDOS) for Hg$_{3}$AsSe$_{4}$I in B, M, D and T structural forms, respectively. Hg$_{3}$AsSe$_{4}$I in both 3D bulk and 2D layered structures exhibit the following electronic characters: their largely dispersed conduction bands are mainly contributed by the hybridized Hg-$6s$, As-$4p$, Se-$4p$, and I-$5p$ orbitals, while their valence band edges are dominated by anion $p$ states. Especially, combining the relativistic effects induced by the heavy Hg and I elements, as well as the net out-of-plane polarization, Rashba SOC effect emerges in 2D vertically polarized Hg$_{3}$AsSe$_{4}$I, splitting their conduction and valence band edges into separated band branches and shifting the positions of valence band maximum (VBM) and conduction band minimum (CBM) away from $\Gamma$ point. As a result, all Hg$_{3}$AsSe$_{4}$I layered structures are semiconductors with indirect band gaps ($E_{g}$). As shown in FIG.~\ref{Figure 2}(b), the semiconducting $E_{g}$ of 1.95, 2.49, 1.94, and 1.51 eV are predicted for Hg$_{3}$AsSe$_{4}$I in B, M, D, and T structural forms, respectively. The significant dependence of $E_{g}$ on system dimension and number of monolayers reveals the strong interlayer electronic coupling within layered Hg$_{3}$AsSe$_{4}$I\cite{001291654100001}. Similar to many other 2D layered materials\cite{000526353000046}, $E_{g}$ of 2D Hg$_{3}$AsSe$_{4}$I layers decreases as a function of monolayer number, yet it remains larger than the thermodynamic equilibrium potential of 1.23 eV required for water-splitting reactions.

Furthermore, the photon energy dependent optical absorption coefficients $\eta_{\omega}$ of Hg$_{3}$AsSe$_{4}$I in B, M, D, and T structural forms are also simulated and displayed in FIG.~\ref{Figure 2}(c). In general, all Hg$_{3}$AsSe$_{4}$I structures under investigation exhibit strong absorption of solar spectrum throughout the visible light range (1.65–3.26 eV). Therefore, the semiconducting 2D Hg$_{3}$AsSe$_{4}$I layers are highly capable for absorption of incident sunlight during the initial step of photocatalytic water-splitting reaction.

\begin{figure*}[!]
\centering
\includegraphics[width=0.92\linewidth]{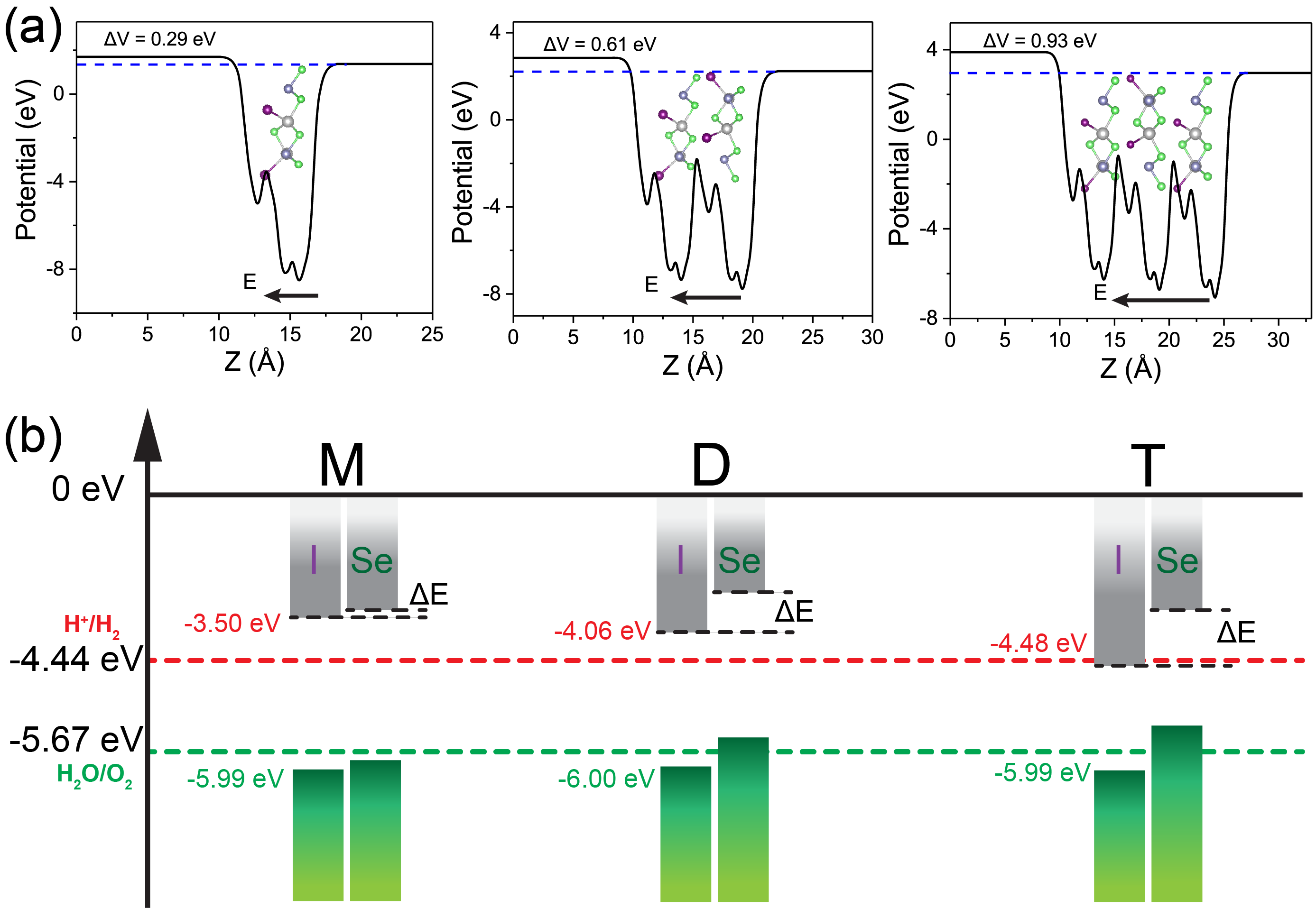}\vspace{-3pt}
\caption{(a) Planar-averaged electrostatic potential profiles for 2D Hg$_{3}$AsSe$_{4}$I in M, D and T structural forms, respectively. The built-in electric field ($\Phi$) orientated from Se to I atomic surfaces are indicated by black arrows. (b) The alignment of band edges from I and Se atomic surfaces for all 2D Hg$_{3}$AsSe$_{4}$I layers. Red and green dash lines indicate the required potentials for the hydrogen evolution reaction (HER) and oxygen evolution reaction (OER), respectively. Owing to the net out-of-plane polarization, the potential offset ($\Delta V$) and energy shift ($\Delta E$) are developed between the top and bottom atomic surfaces of 2D Hg$_{3}$AsSe$_{4}$I layers.}
\label{Figure 3}
\end{figure*}

Besides the absolute band gaps, the alignment of conduction and valence band edges with respect to the water reduction and oxidation potentials are also indispensable for photocatalytic performance. Specifically, the energy levels of CBM and VBM for photocatalysis are required to be higher than the H$^{+}$/H$_{2}$ reduction potential (-4.44 eV) and lower than the H$_{2}$O/O$_{2}$ oxidation potential (-5.67 eV), respectively. In order to determine the relative energy levels of band edges, the plane-averaged electrostatic potentials along the out-of-plane direction for 2D Hg$_{3}$AsSe$_{4}$I layers are simulated and displayed in FIG.~\ref{Figure 3}(a). Owing to net out-of-plane polarization, the electrostatic potential offsets ($\Delta V$) are developed between the Se and I atomic surfaces for all 2D Hg$_{3}$AsSe$_{4}$I layers\cite{000367244000001}. By increasing the number of Hg$_{3}$AsSe$_{4}$I monolayers, $\Delta V$ significantly increases from 0.29 eV of M to 0.93 eV of T. Thus, during the redox reaction, photo-excited $h^{+}$ will preferably transfer to the I surface and $e^{-}$ tend to migrate towards the Se side, respectively. The out-of-plane polarization within 2D Hg$_{3}$AsSe$_{4}$I layers can facilitate the separation of photo-excited carriers and is beneficial to the photocatalytic efficiency. Using the vacuum level as the energy reference ($E$ = 0 eV), the alignment of CBM and VBM energy levels relative to the water reduction and oxidation potentials are presented in FIG.~\ref{Figure 3}(b). Typically, for the Se atomic surfaces with lower electrostatic potential, their band edges shift to higher energy levels relative to the band edges from I atomic surfaces. The thicker Hg$_{3}$AsSe$_{4}$I layers, the larger magnitude of energy shift ($\Delta E$) between the two surfaces are developed. Overall, only 2D Hg$_{3}$AsSe$_{4}$I monolayer has band edges from both atomic surfaces that perfectly satisfy the band alignment requirements for water-splitting reaction. In comparison, Hg$_{3}$AsSe$_{4}$I multilayers either have one of their CBM energy levels lower than the hydrogen evolution reaction (HER) potential or VBM higher than the oxygen evolution reaction (OER) potential. Therefore, as a 2D photocatalyst with nearly ideal band alignment, Hg$_{3}$AsSe$_{4}$I monolayer will be under the detailed simulations and analysis in the remainder of this work.

Driven by the intrinsic built-in electric field, the photo-excited electrons $e^{-}$ and holes $h^{+}$ are spontaneously separated, and further migrate towards their preferable atomic surfaces for HER and OER reactions, respectively. To evaluate the carrier transport properties of Hg$_{3}$AsSe$_{4}$I monolayer, we simulate its carrier mobility $\mu$ based on the Bardeen-Shockley method\cite{A1950UB34000015,A1989U229100027}:

\begin{equation}
\mu_{\beta} = \frac{e\hbar^{3}C_{2D}}{k_{B}Tm_{\beta}^{*}m_{\beta}^{d}(E_{\beta}^{i})^{2}}
\end{equation} 

\noindent where $\beta$ can either be $e^{-}$ or $h^{+}$, $m_{\beta}^{*}$ refers to the carrier effective mass, which can be obtained from parabolic fitting of conduction and valence band edges under the effective mass approximation. The averaged effective carrier mass $m_{\beta}^{d}$ = $\sqrt{m_{x}^{*}m_{y}^{*}}$, where $x$ and $y$ represent the two planar directions under Cartesian coordinates. The deformation potential constant $E_{\beta}^{i}$ = $\partial V_{i}/\partial s$, where $V_{i}$ is the energy level of band edge (CBM for $e^{-}$ and VBM for $h^{+}$) and $s$ is the applied uniaxial strain. 2D elastic modulus $C_{2D}$ = $[\partial^{2}\Delta E/\partial s^{2}]/A_{0}$ equals to the second-order derivative of system energy in response to the uniaxial strain $s$ normalized to monolayer area $A_{0}$. The computational details for simulating $E_{\beta}^{i}$ and $C_{2D}$ are shown in FIG. S3. Using the fitted parameters tabulated in Table~\ref{Table 1}, the carrier mobilities $\mu$ of 2D Hg$_{3}$AsSe$_{4}$I monolayer around room temperature are determined to be: $\mu_{e^{-}}$ = 386.40 cm$^{2}$V$^{-1}$s$^{-1}$ and $\mu_{h^{+}}$ = 98.12 cm$^{2}$V$^{-1}$s$^{-1}$. It is noted that the simulated $\mu$ of Hg$_{3}$AsSe$_{4}$I monolayer is even better than the known 2D photocatalytic materials such as MoS$_{2}$\cite{000335369200029} and PbS$_{2}$\cite{000880841800001}, and is comparable to that of WS$_{2}$\cite{000426233700016}. Therefore, the superior carrier transport properties can be expected in 2D Hg$_{3}$AsSe$_{4}$I monolayer.

\begin{table}[h]
\centering
\caption{The calculated effective masses of $e^{-}$ and $h^{+}$ (in units of $m_{0}$), deformation potential constants (in eV), 2D elastic modulus (in J/m$^{2}$), and carrier mobilities (in cm$^{2}$V$^{-1}$s$^{-1}$) for the Hg$_{3}$AsSe$_{4}$I monolayer.}
\begin{tabular}{ccccccc}
\hline \hline
 $m_{e^{-}}^{*}$    & $m_{h^{+}}^{*}$    & $E_{e^{-}}^{i}$  & $E_{h^{+}}^{i}$     & $C_{2D}$   &   $\mu_{e^{-}}$     &   $\mu_{h^{+}}$  \\
\hline
  0.29                 & 0.52               & 8.91             & 11.07              & 122      &     386.40           &    98.12          \\
\hline\hline
\label{Table 1}
\end{tabular}
\end{table}

\begin{figure*}[tb]
\centering
\includegraphics[width=0.97\linewidth]{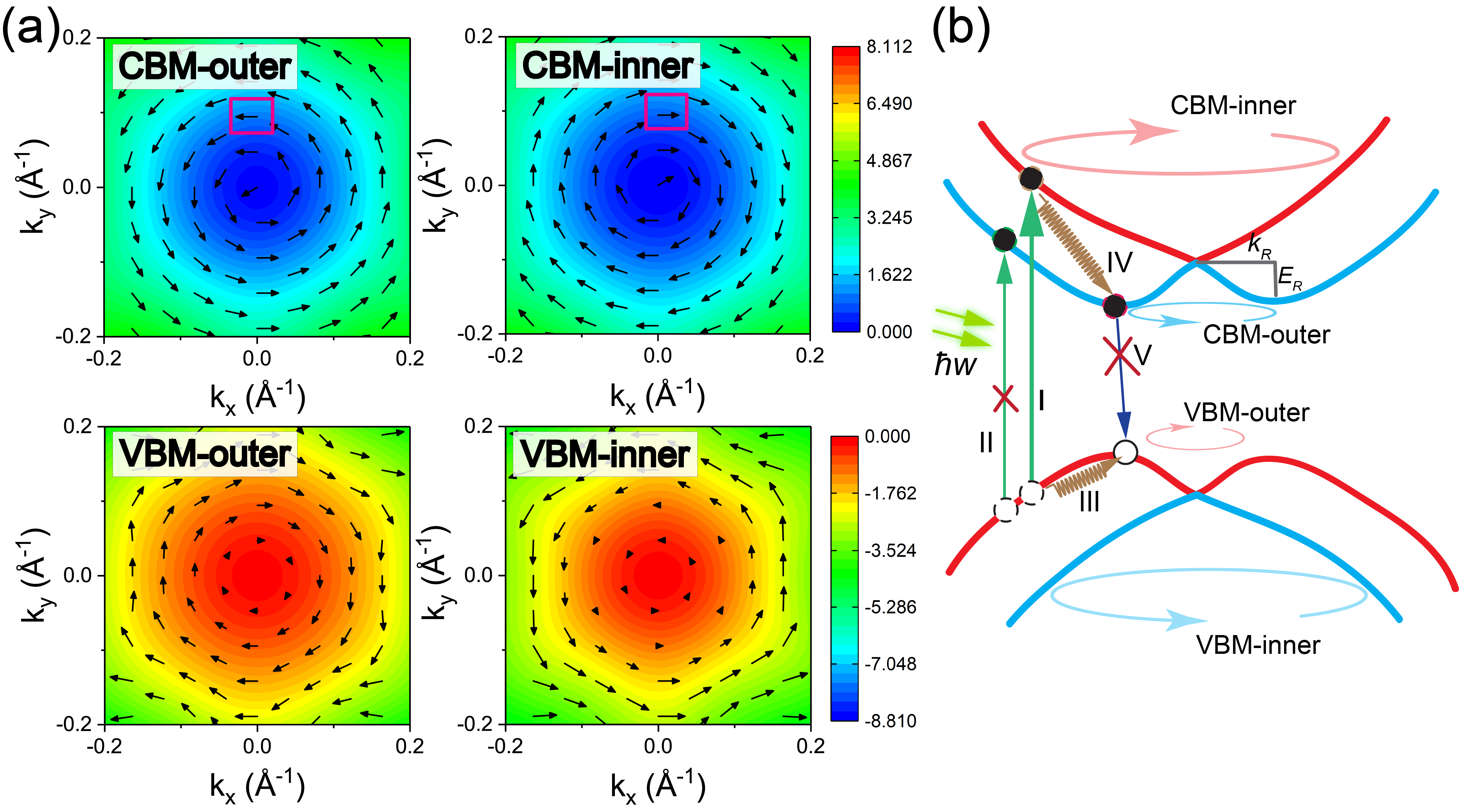}\vspace{-3pt}
\caption{(a) The calculated spin texture plots for Hg$_{3}$AsSe$_{4}$I monolayer among $k_{x} - k_{y}$ reciprocal plane. The spin components of electronic eigenstates are indicated by black arrows. The color bars correspond to the energy eigenvalues relative to the valence and conduction band extrema. VBM-outer (CBM-inner) and VBM-inner (CBM-outer) band branches characterize the ``clockwise'' or ``anticlockwise'' spin orientations, respectively. (b) The schematic illustration of Rashba spin-split band branches as well as carrier migration and recombination pathways. The photo-generated electrons and holes are represented by solid and hollow circles, respectively. The relaxations of carriers to the band edges via inelastic phonon scattering are indicated by the waved arrows. The band branches characterizing ``clockwise'' and ``anticlockwise'' spin orientations are marked in red and blue colors. The recombination of carriers from band branches with different spin orientations are forbidden.}
\label{Figure 4}
\end{figure*}

As we discussed earlier, due to the synergy of vertical polarization and relativistic effects in the form of SOC, Rashba effect lifts the band degeneracy by splitting band edges into separated branches and shifts the positions of CBM and VBM away from $\Gamma$ point. More significantly, Rashba SOC also interconnects the electron momentum and spin degree of freedom, which in turn can prevent photo-excited electrons and holes from recombination. The interconnection between electron momentum $k$ and spin $\sigma$ in Pauli matrix representation within Rashba effect can be formulated as:\cite{000701980400007}
\begin{equation}
H = \alpha \times (k_{y}\sigma_{x}-k_{x}\sigma_{y})\\
\label{eq22}
\end{equation}
\noindent where the Rashba coefficient $\alpha$ can be determined using the momentum and energy offsets around band extrema ($k_{R}$ and $E_{R}$) as: $\alpha  = 2E_{R}/k_{R}$. For Hg$_{3}$AsSe$_{4}$I 3D bulk with a sizable vertical polarization magnitude and significant relativistic effects, $\alpha$ with considerable amplitudes are determined for its valence and conduction band edges ($\alpha_{VBM}$ = 15.40 meV$\cdot${\AA}, $\alpha_{CBM}$ = 135.87 meV$\cdot${\AA}). Based on the formalism associated with Rashba effects given in Eq.~(\ref{eq22}), for a given electronic eigenstate of 2D Hg$_{3}$AsSe$_{4}$I, its spin component $\sigma$ is always perpendicular to electron momentum $k$, leading to the tangential spin texture characters. FIG.~\ref{Figure 4}(a) presents our simulated spin texture plots for individual band branches around VBM and CBM of Hg$_{3}$AsSe$_{4}$I monolayer, by mapping their energy and spin eigenvalues for individual band branch among the $k_{x}-k_{y}$ reciprocal plane. As a fingerprint feature of the Rashba effect, both inner and outer band branches of CBM and VBM have their electron spin components rotating in nearly circular shapes upon the constant energy contours. Specifically, CBM-outer and VBM-inner band branches characterize the ``anticlockwise'' spin orientations, while VBM-outer and CBM-inner band branches have their spin components in ``clockwise'' rotations. 

Based on the spin and energy eigenvalues illustrated by spin texture plots, FIG.~\ref{Figure 4}(b) further displays the schematic diagram of Rashba-split band branches for Hg$_{3}$AsSe$_{4}$I monolayer. Two eigenstates from the CBM(VBM)-outer and -inner band branches with identical energy eigenvalues (indicated by magenta rectangles in FIG.~\ref{Figure 4}(a)) can be related by reversing the sign of electron momentum ($k \rightarrow -k$). Since time-reversal symmetry is preserved under Rashba SOC ($H^{\ast}_{SOC} = H_{SOC}$), the flip of electron spin orientation upon momentum reversal is naturally enforced ($E (-k, \sigma) = E(k, -\sigma)$). Therefore, the inner and outer band branches with ``clockwise'' and ``anticlockwise'' spin orientations indeed have opposite spin components. Considering the spin degree of freedom associated with spin-split band branches of Hg$_{3}$AsSe$_{4}$I monolayer, additional spin-selection rules are involved during the photo-induced carrier excitation and recombination processes\cite{000366339600003,9tfy-gp2t,000701980400007}. As illustrated by FIG.~\ref{Figure 4}(b), under incident sunlight, electrons can be excited from the VBM-outer to CBM-inner branch with the same spin orientations (process I), while the optical excitations between VBM-outer and CBM-outer branches are forbidden (process II). Photo-excited electrons $e^{-}$ and holes $h^{+}$ will be individually relaxed into the band edges via the spin-independent inelastic phonon scattering (processes III and IV)\cite{000614271100003}. As VBM-outer and CBM-outer branches around the band extrema of Hg$_{3}$AsSe$_{4}$I monolayer characterize the opposite spin components, the radiative carrier recombination between VBM-outer and CBM-outer band branches is completely forbidden (process V), leading to the efficient reduction of carrier recombination rate and enhancement of carrier lifetime in 2D Hg$_{3}$AsSe$_{4}$I. Moreover, compared to those ferroelectric semiconducting photocatalysts whose polarizations can be largely reduced at elevated temperature or even completely suppressed by the depolarization field at ultrathin thickness\cite{000417831100007}, the vertical polarization of layered Hg$_{3}$AsSe$_{4}$I is nearly independent of system dimension (see similar spin texture plots for Hg$_{3}$AsSe$_{4}$I bulk in FIG. S4) and is able to survive at room temperature, rendering the additional advantage for practical photocatalytic applications.

\begin{figure*}[t]
\centering
\includegraphics[width=1\linewidth]{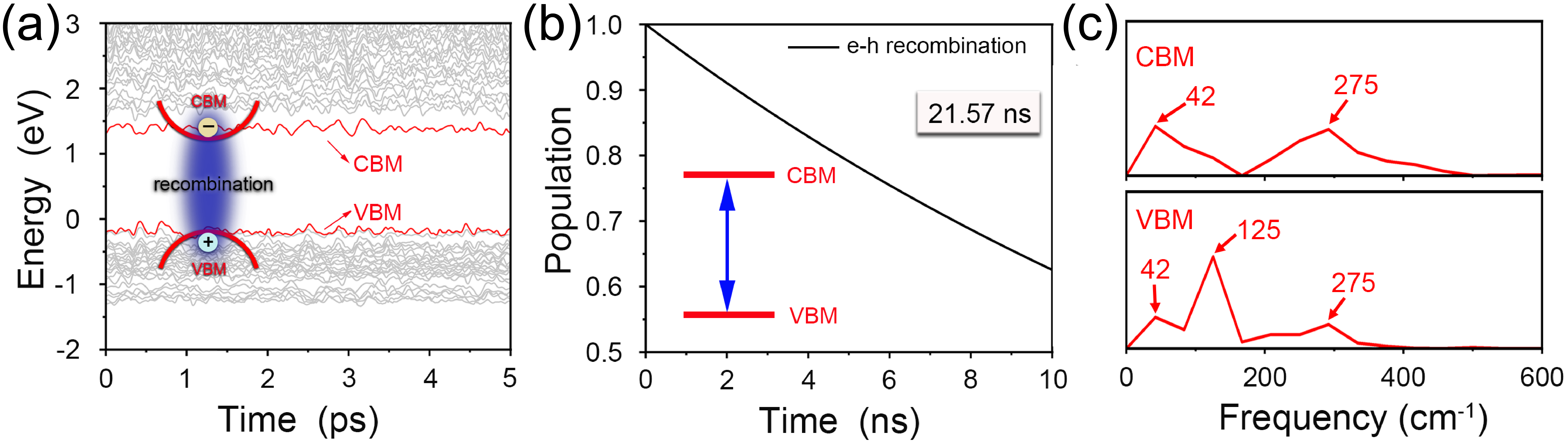}\vspace{-3pt}
\caption{(a) Time dependent evolution of the energy states around band extrema of Hg$_{3}$AsSe$_{4}$I monolayer. CBM and VBM electronic states are indicated by red lines. The inset shows the corresponding electron-hole recombination process between VBM and CBM. (b) Calculated lifetime for the recombination of photogenerated electrons and holes between CBM and VBM electronic states within Hg$_{3}$AsSe$_{4}$I monolayer. (c) Fourier transform of the autocorrelation functions for the CBM and VBM fluctuations in Hg$_{3}$AsSe$_{4}$I monolayer.}
\label{Figure 5}
\end{figure*}

Besides the static electronic structures, the dynamic properties of photo-excited carriers, specifically the time scale associated with the nonradiative electron-hole recombination process within Hg$_{3}$AsSe$_{4}$I monolayer, have also been evaluated using NAMD simulations. During the electron-hole recombination process, the nonadiabatic coupling (NAC) that quantifies the magnitude of inelastic electron-phonon scattering and energy transfer between electrons and phonons, is a primary factor determining the time scale associated with $e^{-}$-$h^{+}$ recombination. Typically, the average NAC element $d_{jk}$ between electronic eigenstates $j$ and $k$ can be formulated as\cite{001097066900004,000935731700001}:

\begin{equation}\label{Eq6}
  d_{jk} = \langle \phi_{j}|\frac{\partial}{\partial t}| \phi_{k} \rangle = \sum_{I} \frac{\langle \phi_{j}|\nabla_{R}H| \phi_{k} \rangle}{\varepsilon_{k}-\varepsilon_{j}}\dot{R}_{I}
\end{equation}

\noindent where $H$ is the Kohn-Sham Hamiltonian, $\dot{R}_{I}$ is the velocity of nuclei, $\phi_{j}$, $\phi_{k}$, $\varepsilon_{k}$, $\varepsilon_{j}$ are the wavefunctions and energy eigenvalues for $j$ and $k$ electronic states, respectively. Clearly, a larger band gap (larger $\varepsilon_{k}-\varepsilon_{j}$ energy difference), weaker electron-phonon coupling strength $\langle \phi_{j}|\nabla_{R}H| \phi_{k} \rangle$, and lower nuclei velocity can lead to an overall small NAC magnitude, which in turn favors the slow nonradiative electron-hole recombination rate and enhanced carrier lifetimes.

FIG.~\ref{Figure 5}(a) displays the time-dependent energy evolutions of electronic states around the band extrema for Hg$_{3}$AsSe$_{4}$I monolayer, obtained from NAMD simulations performed around 300 K. Both VBM and CBM electronic states (marked by red lines in FIG.~\ref{Figure 5}(a)) oscillate within the amplitude of 0.2 eV during the NAMD trajectory. Moreover, driven by the intrinsic out-of-plane polarization, the spatially separated CBM and VBM states result in the overall small wave function overlap and weak contributions to the $\langle \phi_{j}|\nabla_{R}H| \phi_{k} \rangle$ term. Especially, the CBM state is isolated from other low-lying conduction bands (e.g., CBM+1 state), therefore the direct recombination of electrons from CBM and holes from VBM states are extracted for quantitative analysis. The state population of photo-generated carriers between CBM and VBM of Hg$_{3}$AsSe$_{4}$I monolayer is provided in FIG.~\ref{Figure 5}(b). After performing the exponential fitting of time-dependent state population ($P(t) = \exp(-t/\tau)$)\cite{000833894000001}, the time scale associated with nonadiabatic electron-hole recombination is determined to be 21.57 ns. Such a long carrier lifetime within a single-phase vertically polarized 2D monolayer, is comparable to those of 2D heterobilayers with Z-scheme heterojunction and interfacial electric field (e.g., $\tau$ = 11.03 ns for BCN/In$_{2}$S$_{3}$ heterojunction\cite{000413057500080}).

As electron-phonon coupling is crucial for the nonadiabatic electron-hole recombination process, the coupling of phonon modes to electronic states can be revealed by performing Fourier transforms of the time-dependent state energy evolutions (shown in FIG.~\ref{Figure 5}(c)). The peak positions in the Fourier transform spectrum can be assigned to the frequencies of specific phonon modes for Hg$_{3}$AsSe$_{4}$I monolayer. Specifically, electrons from CBM states are strongly coupled to the phonons around 42 and 275 cm$^{-1}$, corresponding to the $E$ and $A_{1}$ vibrational modes of Hg$_{3}$AsSe$_{4}$I monolayer (see FIG. S5 and S6 for illustration). In contrast, holes from VBM strongly couple to $E$ vibrational phonon modes around 125 cm$^{-1}$ (see FIG. S7 for illustration). Owing to the large atomic masses of heavy elements in Hg$_{3}$AsSe$_{4}$I monolayer, phonon modes coupled to photo-generated electrons and holes are in low frequencies, leading to the overall slow nuclei velocity. Therefore, combining the relatively large $E_{g}$, the weak electron-phonon coupling strength, and slow nuclei velocity, the average NAC value between CBM and VBM states is as small as 0.68 meV. The small NAC value is responsible for slow carrier recombination and long carrier lifetime. As a result, photo-generated electrons and holes with long lifetimes can efficiently participate in HER and OER processes within Hg$_{3}$AsSe$_{4}$I monolayer, which is highly beneficial for optimized photocatalytic performance.

\section{\label{sec:level1}IV.CONCLUSION}

In the current work, the photocatalytic water-splitting performance of experimentally synthesizable 2D vertically polarized Hg$_{3}$AsSe$_{4}$I has been systematically investigated by simulating its electronic structure, optical absorption, and carrier recombination properties based first-principles calculations and time-domain NAMD simulations. Hg$_{3}$AsSe$_{4}$I configurations in both 3D bulk and 2D layered structural forms characterize the semiconducting $E_{g}$ and strong absorption of incident photons across the visible-light spectrum. However, only the Hg$_{3}$AsSe$_{4}$I monolayer possesses valence and conduction band edges that can perfectly meet the band alignment required for water reduction and oxidation potentials, fulfilling the fundamental optical and electronic prerequisites for light-driven water-splitting reactions. The optimized carrier transport properties and considerable carrier mobilities ($\mu_{e^{-}}$ = 386.40 cm$^{2}$V$^{-1}$s$^{-1}$ and $\mu_{h^{+}}$ = 98.12 cm$^{2}$V$^{-1}$s$^{-1}$) are also predicted for Hg$_{3}$AsSe$_{4}$I monolayer, outperforming the most known 2D photocatalysts. Owing to the asymmetric anion distributions between top and bottom atomic surfaces of 2D Hg$_{3}$AsSe$_{4}$I, the stable and irreversible out-of-plane polarization that can overcome the critical thickness limitation arises inside the Hg$_{3}$AsSe$_{4}$I monolayer, rendering a built-in electric field that can effectively separate photo-excited electrons and holes from recombination and facilitate the sequential photocatalytic hydrogen and oxygen production. More importantly, combining the out-of-plane polarization and strong relativistic effects induced by heavy elements (e.g., Hg and I), SOC effect within the Rashba formalism emerges in Hg$_{3}$AsSe$_{4}$I, splitting the valence and conduction band edges into spin-resolved band branches with unique spin-momentum locking characters, which can effectively introduce the spin-selection rules that forbid the direct radiative recombination of photo-excited electron-hole pairs. The synergistic effect of intrinsic vertical polarization and Rashba SOC creates the dual mechanism for significant improvement of carrier separation efficiency and carrier lifetime within Hg$_{3}$AsSe$_{4}$I monolayer. Time-domain NAMD simulations further predict an ultra-long photo-excited carrier lifetime up to 21.57 ns for Hg$_{3}$AsSe$_{4}$I monolayer. The ultra-long carrier lifetime originates from the weak nonadiabatic electron-phonon coupling strength, ensuring the sufficient time scale for photo-excited carriers to participate in water-splitting reactions. Our work not only identifies the Hg$_{3}$AsSe$_{4}$I monolayer as an ideal 2D photocatalyst for water-splitting reactions, but also demonstrates a rational strategy for simultaneously improving the photocatalytic efficiency and prolonging the carrier lifetime via the synergistic mechanism associated with out-of-plane polarization and Rashba SOC effect in 2D vertically polarized materials.

\section{\label{sec:level1}ACKNOWLEDGMENTS}
\begin{acknowledgments}
Authors acknowledge the funding support from the National Natural Science Foundation of China (Grant No. 11574244), and State Key Laboratory of Electrical Insulation and Power Equipment (Grant No. EIPE26318). Hefei Advanced Computing Center is acknowledged for computational support.
\end{acknowledgments}

\section{\label{sec:level1}DATA AVAILABILITY}
The data are available from the authors upon reasonable request.

\bibliography{reference}
\end{document}


\title{\textbf{Supplemental Material for ``Two-dimensional vertically polarized Hg$_{3}$AsSe$_{4}$I monolayer for efficient photocatalytic water-splitting: promoting carrier separation by intrinsic electric field and Rashba effect''} 
}%

\author{Xinfeng Chen}
\altaffiliation{These authors contributed equally to this work.}
\affiliation{Frontier Institute of Science and Technology, State Key Laboratory of Electrical Insulation and Power Equipment, Xi’an Jiaotong University, Xi’ an 710049, People's Republic of China}
\affiliation{Key Lab of advanced optoelectronic quantum architecture and measurement (MOE), Beijing Key Laboratory of Quantum Matter State Control and Ultra-Precision Measurement Technology, and School of Physics, Beijing Institute of Technology, Beijing 100081, China}

\author{Wenchao Shan}
\altaffiliation{These authors contributed equally to this work.}

\author{Fengfeng Ye}

\author{Gaoyang Gou}
\email{gougaoyang@mail.xjtu.edu.cn}
\affiliation{Frontier Institute of Science and Technology, State Key Laboratory of Electrical Insulation and Power Equipment, Xi’an Jiaotong University, Xi’ an 710049, People's Republic of China}

\maketitle

\newpage

\begin{figure}[h]
\centering
\includegraphics[width=0.98\linewidth]{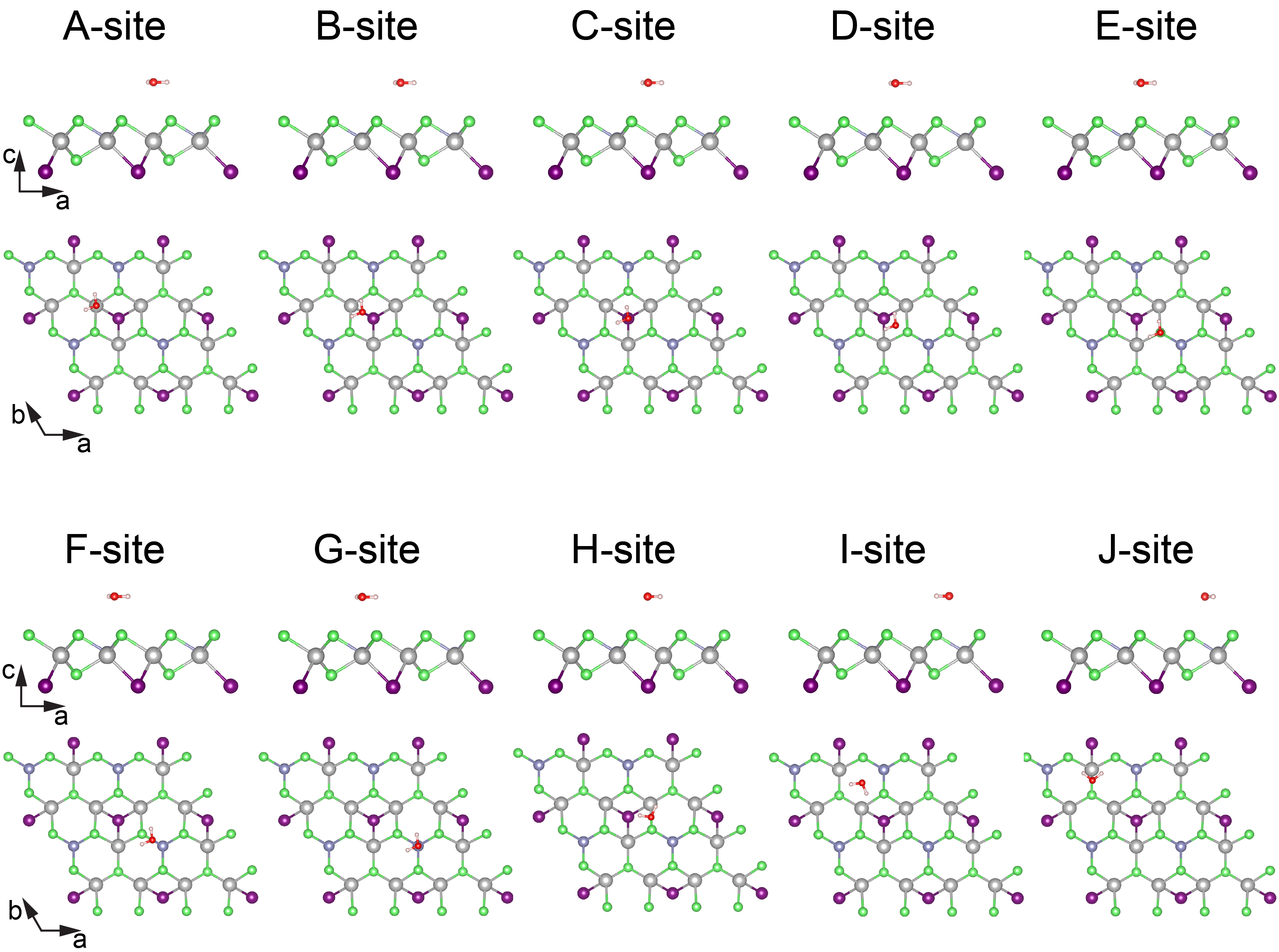}
\renewcommand{\figurename}{FIG. S}
\caption{All possible configurations for adsorption of one H$_{2}$O molecule on the Se atomic surface of Hg$_{3}$AsSe$_{4}$I monolayer in 2 $\times$ 2 $\times$ 1 supercell. C-site is determined to be the most preferable adsorption configuration at Se atomic surface.}
\label{FigS1}
\end{figure}

\begin{figure}[h]
\centering
\includegraphics[width=0.98\linewidth]{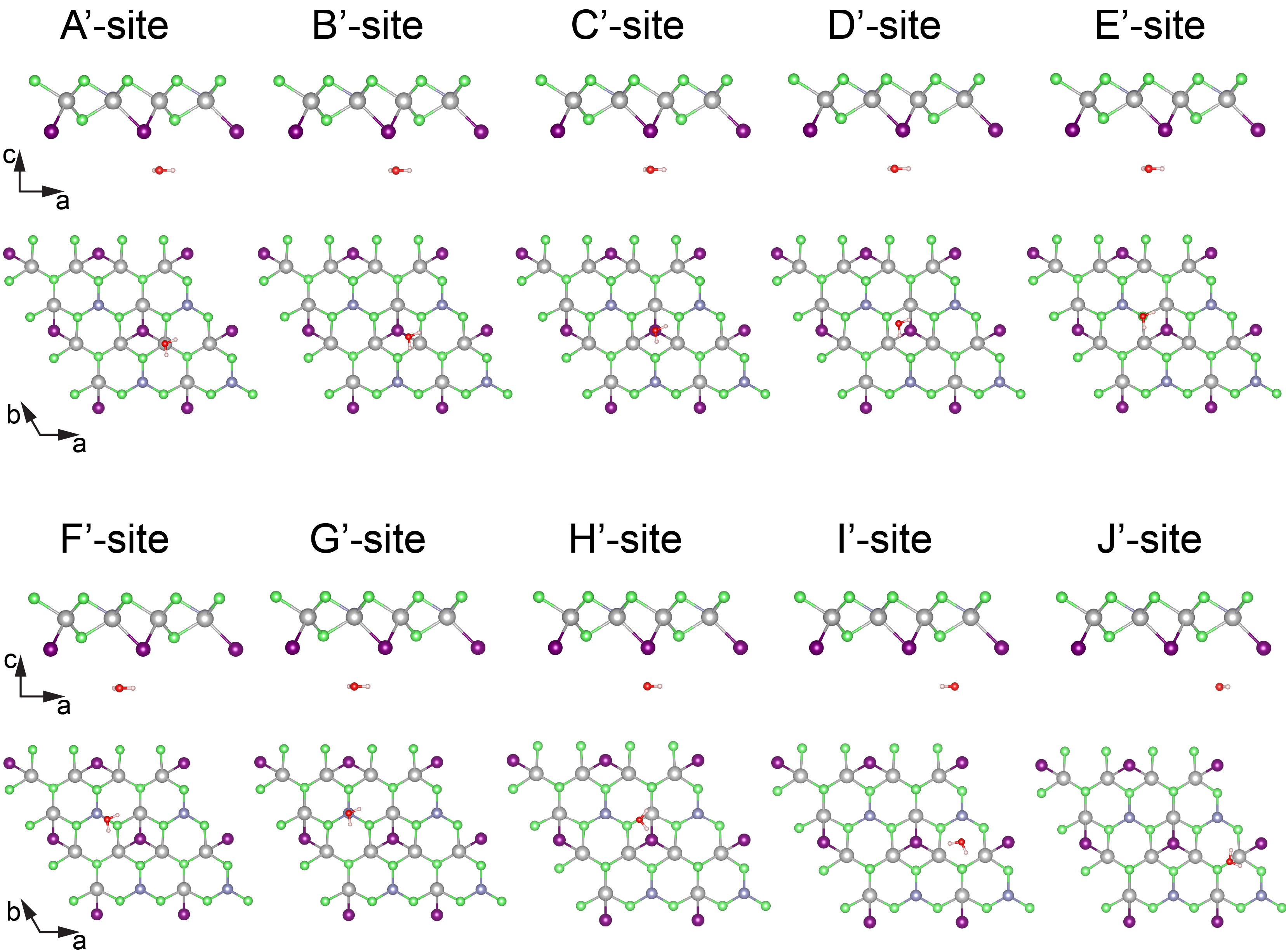}
\renewcommand{\figurename}{FIG. S}
\caption{All possible configurations for adsorption of one H$_{2}$O molecule on the I atomic surface of Hg$_{3}$AsSe$_{4}$I monolayer in 2 $\times$ 2 $\times$ 1 supercell. I$^{'}$-site is determined to be the most preferableadsorption configuration at I atomic surface.}
\label{FigS2}
\end{figure}

\clearpage

\begin{figure}[t]
\centering
\includegraphics[width=0.82\linewidth]{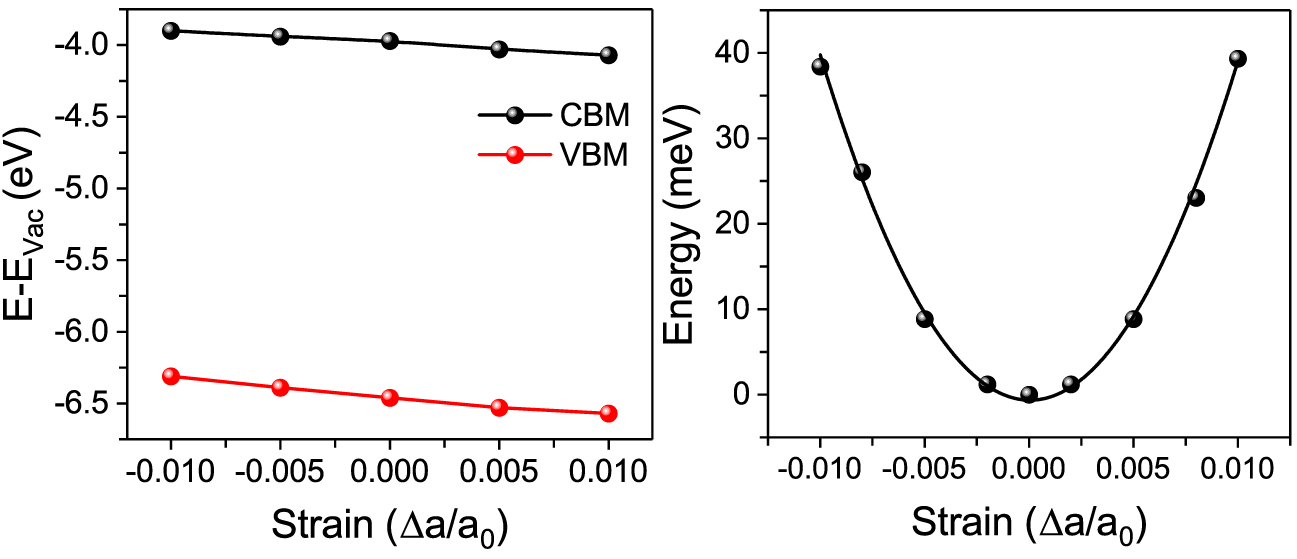}\vspace{-5pt}
\renewcommand{\figurename}{FIG. S}
\caption{Evolution of band edge positions and relative energy $\Delta E$ of Hg$_{3}$AsSe$_{4}$I monolayer as a function of the uniaxial strain $s$ applied along the planar $a$ axis. $E_{\beta}^{i}$ and $C_{2D}$ can be obtained after polynomial fitting of the corresponding curves.}
\label{FigS3}
\end{figure}

\begin{figure}[b]
\centering
\includegraphics[width=0.75\linewidth]{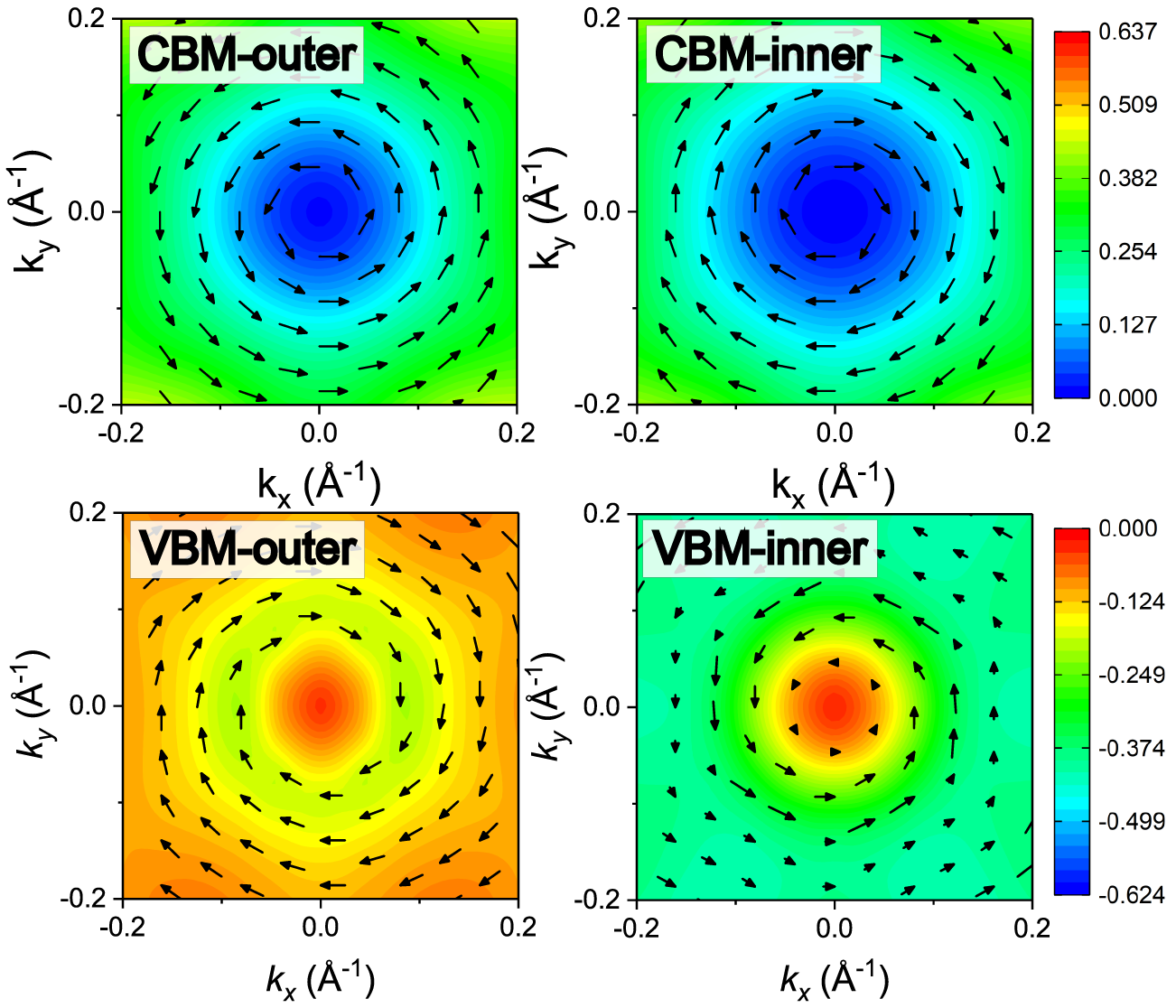}\vspace{-5pt}
\renewcommand{\figurename}{FIG. S}
\caption{The simulated spin texture plots among $k_{x}$$-$$k_{y}$ reciprocal plane for CBM-inner/-outer and VBM-inner/-outer band branches of 3D Hg$_{3}$AsSe$_{4}$I bulk. The spin components of given electronic eigenstates are indicated by black arrows. The color bars represent the energy eigenvalues relative to the valence and conduction band extrema.}
\label{FigS4}
\end{figure}

\clearpage

\begin{figure}[t]
\centering
\includegraphics[width=0.85\linewidth]{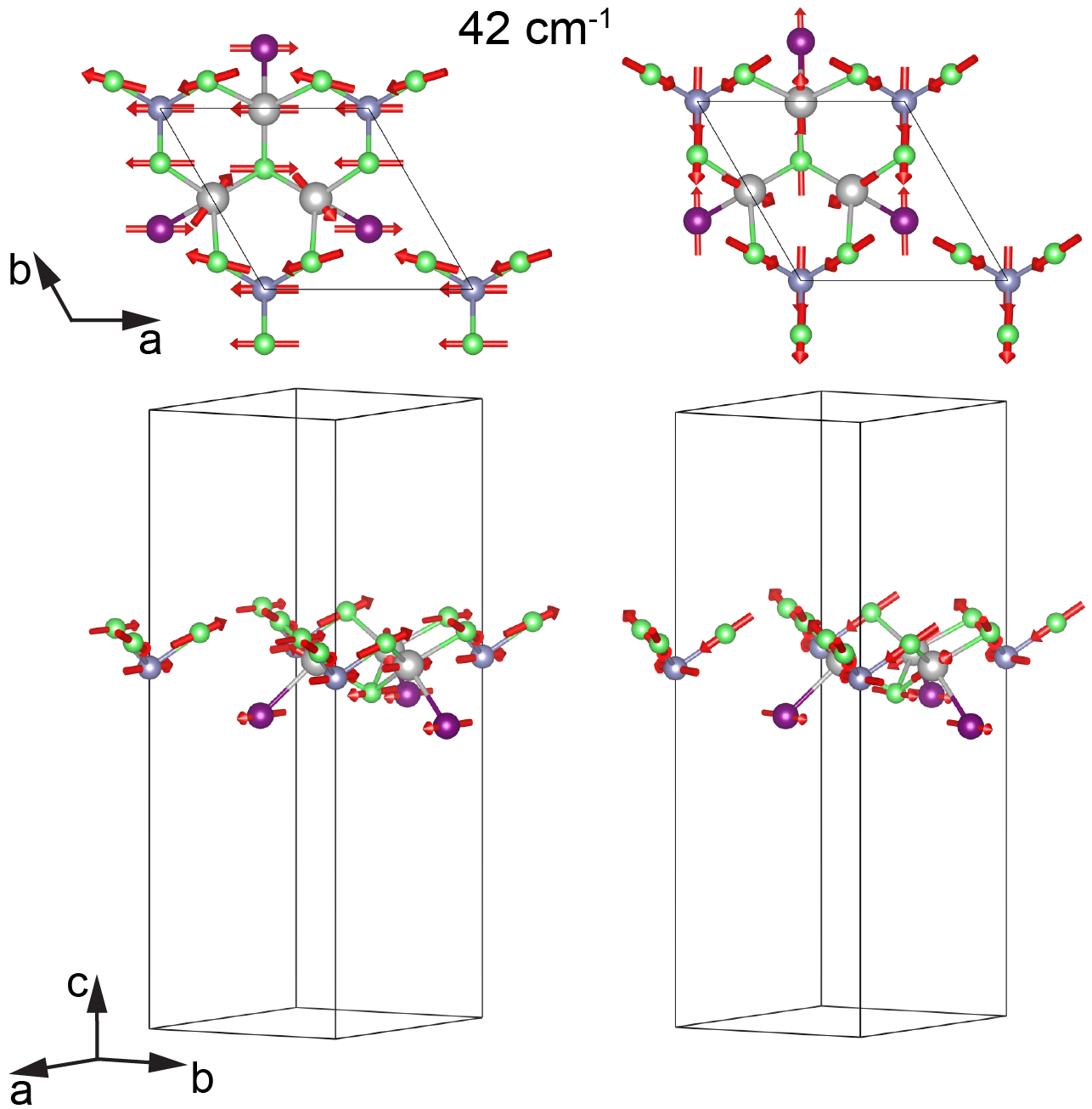}
\renewcommand{\figurename}{FIG. S}
\caption{The two-fold degenerate $E$ phonon modes around 42 cm$^{-1}$ are strongly coupled to the electrons from CBM of Hg$_{3}$AsSe$_{4}$I monolayer. Red arrows indicate the atomic vibrational eigenvectors.}
\label{FigS5}
\end{figure}

\begin{figure}[b]
\centering
\includegraphics[width=0.6\linewidth]{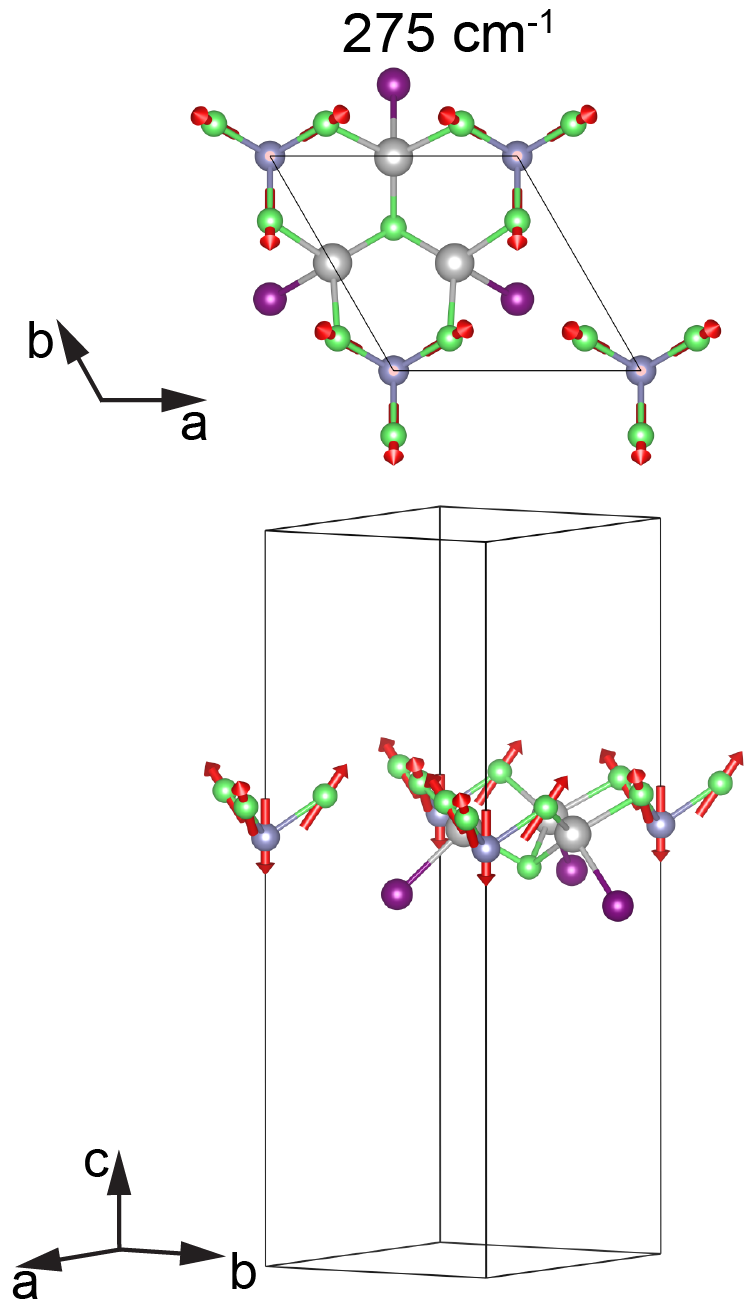}
\renewcommand{\figurename}{FIG. S}
\caption{The $A_{1}$ phonon mode around 275 cm$^{-1}$ is in strong coupling with the electrons from CBM states of Hg$_{3}$AsSe$_{4}$I monolayer.}
\label{FigS6}
\end{figure}

\begin{figure}[b]
\centering
\includegraphics[width=0.85\linewidth]{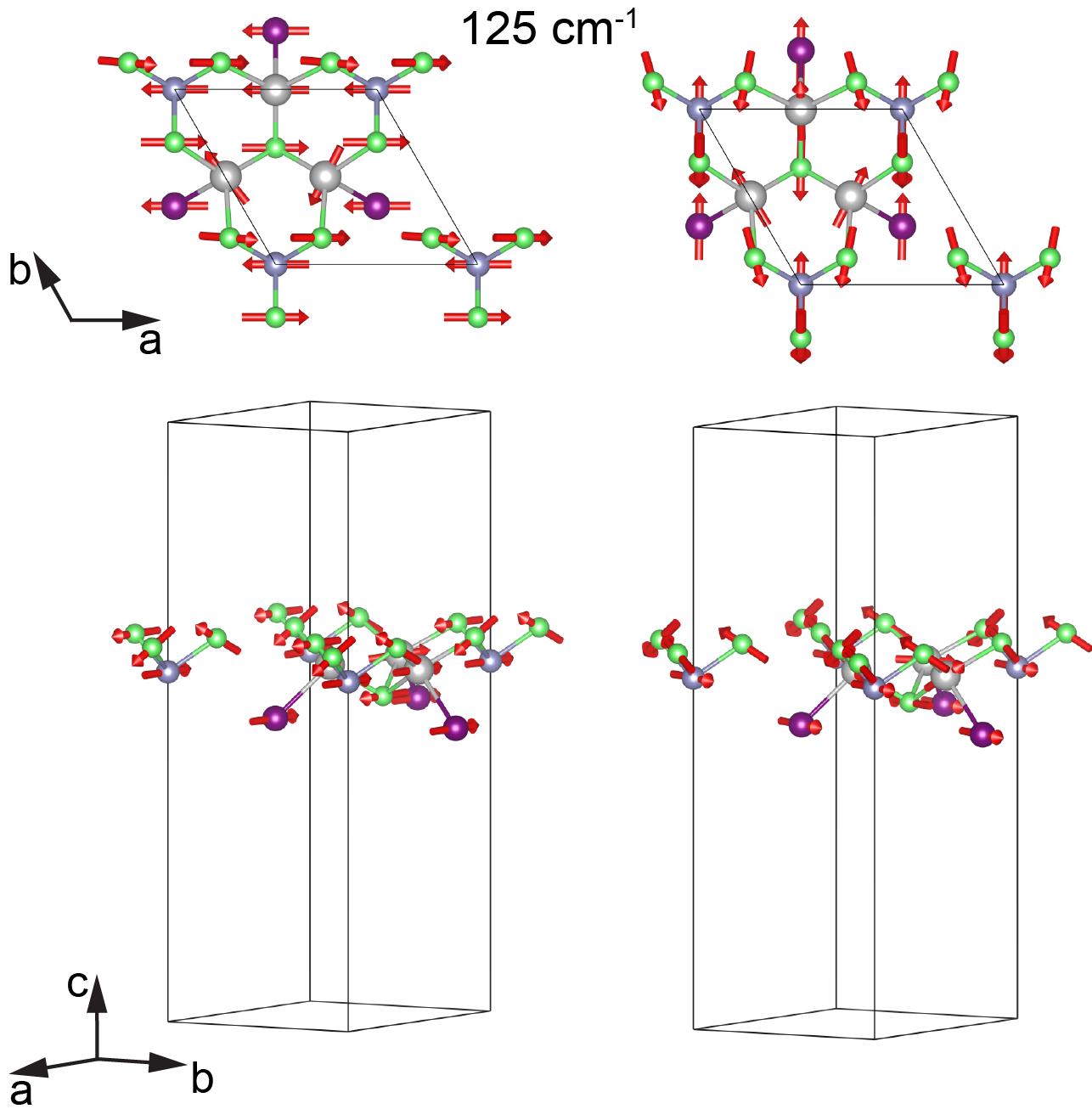}
\renewcommand{\figurename}{FIG. S}
\caption{The two-fold degenerate $E$ phonon modes around 125 cm$^{-1}$ strongly couple to the holes of Hg$_{3}$AsSe$_{4}$I monolayer.}
\label{FigS7}
\end{figure}

\clearpage